# Invisibility on demand based on a generalized Hilbert transform

Zeki Hayran[a], Ramon Herrero[b], Muriel Botey[b], Hamza Kurt[a], and Kestutis Staliunas[b,c]

[a]Nanophotonics Research Laboratory, TOBB University of Economics and Technology, Department of Electrical and Electronics Engineering, Ankara, Turkey
[b]Departament de Física, Universitat Politècnica de Catalunya (UPC), Barcelona, Spain
[c]Institució Catalana de Recercai Estudis Avançats (ICREA), Barcelona, Spain
*Correspondence and requests for materials should be addressed to Z.H. (email: zekihayan@etu.edu.tr).*

**Designing invisible objects without the usage of extreme materials is a long-sought goal for photonic applications. Invisibility techniques demonstrated so far typically require high anisotropy, gain and losses, while also not being flexible. Here we propose an invisibility approach to suppress the scattering of waves from/to given directions and for particular frequency ranges, *i.e.* invisibility on demand. We derive a Born approximation-based generalized Hilbert transform for a specific invisibility arrangement relating the two quadratures of the complex permittivity of an object. The theoretical proposal is confirmed by numerical calculations, indicating that near-perfect invisibility can be attained for arbitrary objects with low-index contrast. We further demonstrate the cases where the idea can be extended to high-index objects or restricted to within practical limits by avoiding gain areas. The proposed concept opens a new route for the practical implementation of complex-shaped objects with arbitrarily suppressed scatterings determined on demand.**

Full invisibility, or cloaking, was proposed using transformation optics or, equivalently, conformal mapping[1,2]. The idea is elegant and fascinating; however, it can hardly cross the limits of science fiction, since the complexity of the required metamaterials severely limit practical realizations. Therefore, actual cloaking schemes generally scarify the perfect wavefront reconstruction or operate under a narrow bandwidth, as for instance in carpet cloaking[3-7], plasmonic cloaking[8,9], or mantle cloaking with thin patterned metasurfaces[10], metallic scatterer[11] or dielectric coating[12] based cloaking, among others[13,14].

A completely different approach to the concept of invisibility, referred as "unidirectional invisibility", relays on systems described by non-Hermitian Hamiltonians[15-17]. The concept is based on the property of an object to be invisible when probed by a wave from one side only. Such effect is accomplished by specific complex-modulated potentials (in optical terms: specific refraction index and gain/loss distributions), that allow suppressing the scattering of radiation from an object.

Unidirectional invisibility was first proposed for parity-time (PT) symmetric periodical systems (defined by symmetric index modulations accompanied by anti-symmetric gain/loss distributions), close to so-called PT-symmetry breaking point. Initially proposed for narrow frequency bands (due to the resonances of the periodic structure), and for particular incidence directions[18-20], the idea was extended to broad band radiation (both in frequency and in propagation direction) also by considering non-PT-symmetric potentials[21-25].

More recently, "unidirectional invisibility" has been related to the more general class of non-Hermitian potentials fulfilling the spatial Kramers-Kronig (KK) relations[26]. In the same way as the causality in time imposes KK relations in frequency, analogously, the KK theory may be directly extended to attain unidirectional invisibility in space. Yet temporal causality implies "invisibility of the future", so the response function $\chi(t)$, which is the kernel in the integral expression of the response of any physical system $A_{\text{resp}}(t) = \int A_{\text{sign}}(t - t_1)\chi(t)\mathrm{d}t_1$, must be zero for all $t > 0$. Such response function in time domain, $\chi(t)$, being $\chi(t) = 0$ for $t > 0$, determines the integral relations between the real and imaginary parts of its spectrum, $\chi_{\text{im}}(\omega) = \frac{1}{\pi} P \int_{-\infty}^{\infty} \frac{\chi_{\text{re}}(\omega_1)}{(\omega-\omega_1)} d\omega_1$, and $\chi_{\text{re}}(\omega) = \frac{-1}{\pi} P \int_{-\infty}^{\infty} \frac{\chi_{\text{im}}(\omega_1)}{(\omega-\omega_1)} d\omega_1$, where $P$ means the Cauchy principle value of the integral. This KK relation (or more generally Hilbert transform) in frequency domain can be directly rewritten in space domain. The spatial invisibility requires that the scattering function fulfills $\chi(\mathbf{k}) = 0$ for all $\mathbf{k} = (k_x, k_y, k_z)$ with $k_x < 0$, i.e. it vanishes on the entire left half-space in wavevector domain. Throughout the letter, the invisibility on demand is formulated in two spatial dimensions (2D), and may be straightforwardly extended to 3D. Such a condition in the 2D **k**-space, $(k_x, k_y)$, leads to similar KK relations in space domain: $\chi_{\text{im}}(x, y) = \frac{1}{\pi} P \int_{-\infty}^{\infty} \frac{\chi_{\text{re}}(x_1, y)}{(x-x_1)} dx_1$, and $\chi_{\text{re}}(x, y) = \frac{-1}{\pi} P \int_{-\infty}^{\infty} \frac{\chi_{\text{im}}(x_1, y)}{(x-x_1)} dx_1$. Therefore, the spatial KK relations are at the basis of unidirectional invisibility: all plane waves propagating from left to right $k_x > 0$, will not be back-scattered by such potentials, since they are uncoupled from the waves propagating to the left $k_x < 0$, as illustrated in Fig. 1a,b. Following the spatial KK approach, a limited bidirectional invisibility has also been proposed, however restricted to small incident angles onto the interface (grazing incidence), by manipulating the reference point of the wavevectors cut-off[24]; bidirectional invisibility was also previously predicted in PT-symmetric systems for a given number of unit-cells[25].

Another class of materials exhibiting reflectionless phenomena have also been so far demonstrated using metasurfaces[27-30] and photonic crystals[31,32]. The complete elimination of the back-scatterings in metasurfaces relies on electric and magnetic Mie-type mode supporting nanostructures, whereas photonic crystal based reflectionless property has been achieved via impedance matching of multiple resonant Bloch eigen-modes and background propagating modes. However to enable the required resonance behavior, materials with sufficiently high refractive indices have to be used[31,33]. It may be highly desirable to extend the reflectionless property to a much broader range of materials (including high-index materials as well as moderate- and low-index materials), to fully exploit the nanofabrication capabilities enabled by emerging industry-compatible technologies[34].

We propose here invisibility on demand: we derive an explicit integral relation, which may be regarded as a generalized Hilbert transform, associated with an arbitrary area of invisibility in **k**-space, see the schematic illustration on Fig. 1c,d. In other words, we propose a procedure to modify the scattering from an object, being either a refractive index scatterer or a complex scatterer including index and gain/loss profiles, in such a way that the object becomes invisible for a range of frequencies and illumination/detection arrangement, depending on the required situation.

The main motivation for the invisibility on demand is facing invisibility under a realistic scope. First, generally full unidirectional invisibility may not be needed in many situations. Moreover, unidirectional invisibility requires a rather severe modification of the potential: if the refraction index - proportional to real part of susceptibility function of the potential, $\chi_{\text{re}}(x, y)$ - is modulated with a specific amplitude, then the required profile of the imaginary part of susceptibility $\chi_{\text{im}}(x, y)$ to render the object invisible, is of the same order of magnitude, as follows directly from the spatial KK relations. In our case, for invisibility on demand, the modification of the complex optical potential, depends upon the area of invisibility in **k**-space. Thus, the proposed scheme of invisibility on demand, working solely for special angular and frequency ranges is substantially more feasible, as shown in details below. Besides, for unidirectional invisibility as based on spatial KK relations the imaginary part of susceptibility decays weakly, as $1/|x|$; which may result inadequate for applications since, as a consequence, the norm of the total gain/loss modification function diverges logarithmically. We show that invisibility under demand, in particular for smooth invisibility boundaries, leads to more convenient asymptotic behaviors (an exponential decay in space, results in a finite norm).

In the letter, we first derive the generalized Hilbert transform, for an arbitrary area of invisibility in wavevector domain. Then, we provide a series of numerical finite-difference-time-domain (FDTD) simulations of different cases to prove the idea and illustrate its performance. Finally, we describe an iterative procedure to generate the complex potentials for invisibility on demand with additional restrictions, for instance avoiding the areas of gain or negative index materials.

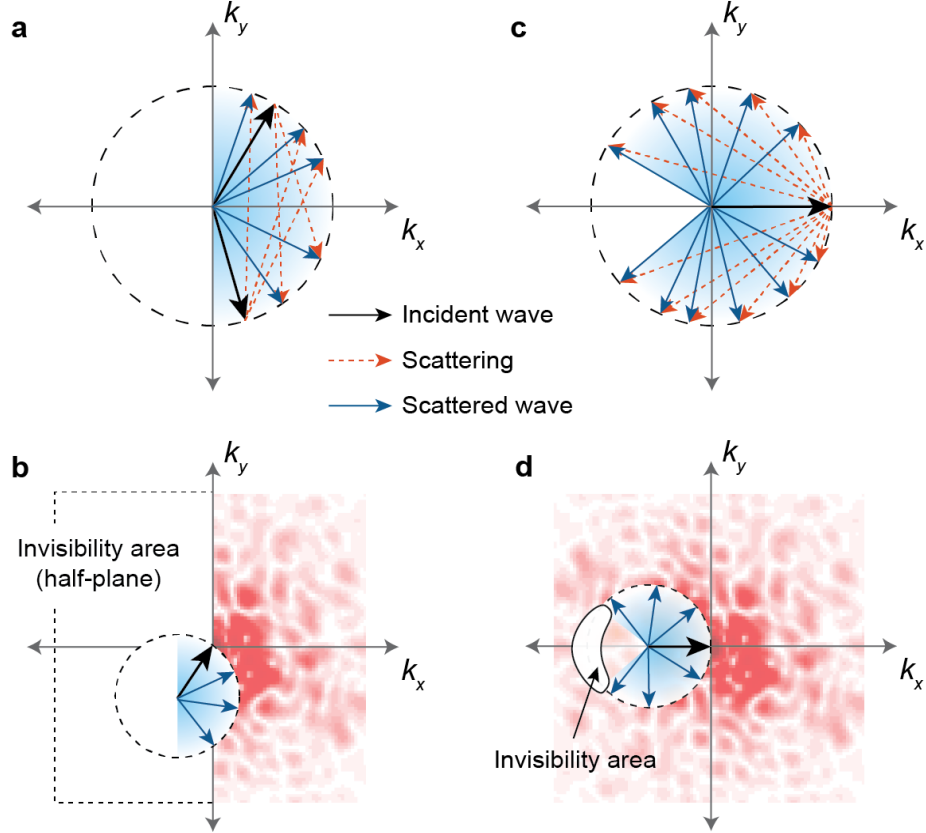

**Figure 1 | Illustration of unidirectional invisibility and invisibility on demand.** (**a,b**) In full unidirectional invisibility, in order to prevent left-reflection of every right-propagating wave, (**b**) all modulation components on the left half-plane, $k_x<0$, in $\mathbf{k} = (k_x, k_y)$ space, must be set to zero. **c,d**, For invisibility on demand, to prevent back-scattering in a particular angular range (**d**) the modulation components to be set to zero are just the ones within a limited invisibility area in $\mathbf{k} = (k_x, k_y)$ space. In particular, the invisibility case (**c**) requires uncoupling the scattering of incident waves around $\mathbf{k} = (k_0, 0)$ into back reflections, i.e. into waves in the vicinity of $\mathbf{k} = (-k_0, 0)$; then, the invisibility function has to be centered at $\mathbf{k} = (-2k_0, 0)$.

## Results

**Generalized Hilbert transform.** Let us consider scattering from a local modulation of the optical potential characterized by the electric permittivity profile $\chi_{\text{re}}(x,y) = \chi_{\text{background}} + a(x,y)$. (Local modulation means $a(x,y) \to 0$, for $(x,y) \to \infty$.) The first order scattering form factor $a(k_x, k_y)$ is simply the Fourier transform of $a(x,y)$: $a(k_x, k_y) = \frac{1}{2\pi} \int a(x,y) \exp(ik_x x + ik_y y) \mathrm{d}x \mathrm{d}y$ (for weak potentials, in the so-called Born approximation, secondary scattering can be neglected). Plane waves with carrier wave-vector $\mathbf{k}_0$ are scattered by each wavenumber component $\mathbf{k}$, of the modulation of the potential, into $\mathbf{k}_1 = \mathbf{k} + \mathbf{k}_0$. Therefore, preventing scattering for a given angular range and in a restricted frequency range, requires imposing that a specific area in $\mathbf{k}$-space leads to no scattering, see Fig. 1d for illustration. Let us define the invisibility function $\theta(k_x, k_y)$ in such a way that $\theta(k_x, k_y) = 1$ in the given area of invisibility, being $\theta(k_x, k_y) = 0$ elsewhere, as shown in Fig. 2a. Next, we construct a new scattering function of the object as: $a_1(k_x, k_y) = a(k_x, k_y) - a(k_x, k_y)\theta(k_x, k_y)$, which only eliminates the scattering from the particular invisibility area in **k**-space. In spatial domain, this corresponds to the modification of the profile of the (complex) susceptibility: $a_1(x, y) = a(x, y) - \frac{1}{2\pi} \iint a(x_1, y_1)\theta(x - x_1, y - y_1) dx_1 dy_1$, since the multiplication of functions in wavenumber domain results in convolution in space. Then, it merely remains to calculate the kernel of the convolution, as the inverse Fourier transform of the invisibility area, $\theta(k_x, k_y)$, and to compute the above convolution. This results in what may be regarded as a generalized Hilbert transform.

In particular, the elimination of scattering from the entire left half-plane, i.e. $\theta(k_x, k_y) = 0$ for all $k_x < 0$, results in $\theta(x, y) = \frac{-i}{\sqrt{2\pi}x}$, and the corresponding convolution leads to the conventional KK or Hilbert transform in space. We thus generalize the Hilbert transform for arbitrary area of invisibility, which results in a different kernel.

Only specific shapes of the invisibility area in **k**-space allow analytical expressions for the kernel of the convolution $\theta(x, y)$. For instance, circular or elliptical invisibility areas result in Bessel functions in direct space $(x, y)$ while square or rectangular areas of invisibility lead to sinc-shaped kernel functions. Some of such analytical cases are presented in the Supplementary Note 1. Especially interesting is the case of a ring-shaped invisibility, or a "partial-sun-eclipse" invisibility area, since it allows full unidirectional invisibility from a monochromatic plane probe wave. These two latter cases also allow analytical kernels, as derived in Supplementary Note 1.

Invisibility on demand may also be realized for smooth invisibility functions. For instance, if the considered invisibility function has Gaussian profile centered around some $-2\mathbf{k}_0$ point: $\theta(k_x, k_y) = exp(-(\mathbf{k} + 2\mathbf{k}_0)^2/\Delta k^2)$, then its inverse Fourier transform is also Gaussian, and leads to a convenient convolution kernel $\theta(x, y)$, with rapidly decreasing asymptotics. Figure 2b,c shows the kernel for an elliptic Gaussian profile (Fig. 2a) to suppress reflections of light. The Gaussian case is analyzed in Supplementary Note 2.

As an example of a possible realization of invisibility on demand, we consider a circular object (a disc, Fig. 2d) to be rendered invisible with respect to an elliptical invisibility area shown in Fig. 2a. The scattering from the disc is a Bessel function in wavevector space as depicted in Fig. 2e. The modified object to be invisible under reflections (Fig. 2g,j) is simply calculated as the product of the object and the kernel spectra (Fig. 2f). The above described method can be performed with respect to any invisibility area in wavenumber domain, and any object.

Besides, we may apply an iterative approach that allows regularizing the complex scattering function with additional constrains, for instance excluding gain or negative index materials, yet allowing losses. In principle, the proposed invisibility scheme would typically lead to media with optical gain, and technical inconveniences might arise either for the fabrication of actual systems (within the conventional nanophotonics). Therefore, to avoid such difficulties, we apply a chain of Hilbert transforms where, at each iteration, negative imaginary parts of the complex susceptibility are completely eliminated by setting gain to zero. This operation is iteratively performed until the negative imaginary part is sufficiently reduced (its convergence depending on the initial refractive index, the size and shape of the invisibility hole).

Such procedure is illustrated in Fig. 2, where the final invisible object only presenting refraction index modulation, and the profile of losses is calculated following the iterative process (the first iteration is in Fig. 2g,j; second in Fig. 2h,k and third in Fig. 2i,l). The invisibility area and corresponding real and imaginary parts of the kernel are shown in Fig. 2a,b,c, respectively. The invisible object in **k**-space of Fig. 2f exhibits a shadowed region as compared to visible object, shown in Fig. 2d,e in direct and reciprocal space, respectively. It is interesting to remark that kernels with smooth profiles in **k**-space, introduce relatively small gain profiles in the modified object. On the contrary, stepwise profiles generate objects with symmetric gain and loss areas that, which however, can be easily converted into only loss areas using the iteration process described above. Besides, interestingly, invisibility areas with even functions in **k**-space directly result, following the Hilbert transform procedure, in purely real scattering functions; therefore, in this case, no iterations are required as demonstrated in Supplementary Note 3.

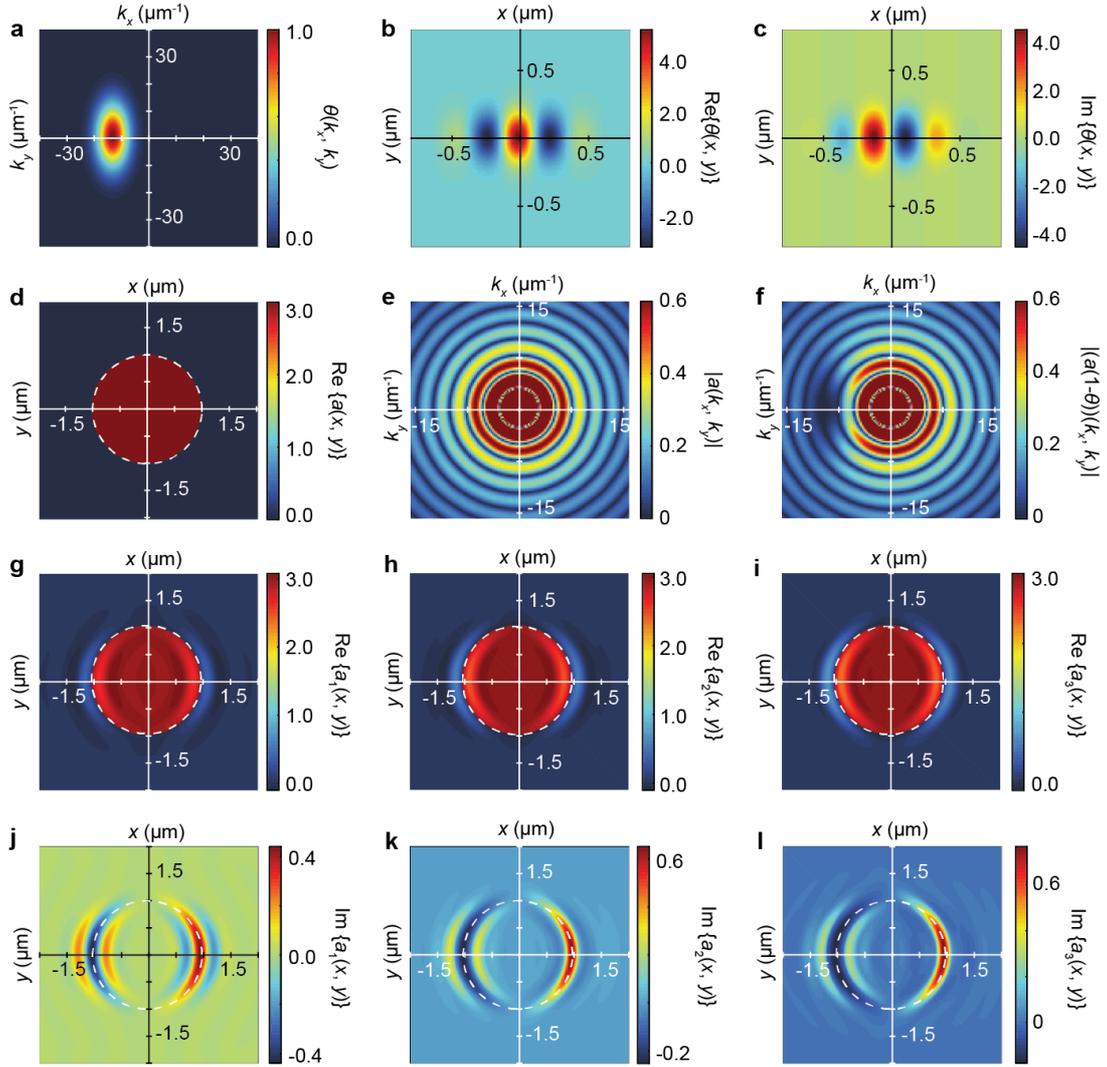

**Figure 2 | Invisibility on demand procedure workflow in spatial and wavevector domain.** (**a-c**) Area of invisibility in (**a**) **k**-space: $\theta(k_x, k_y)$ and the corresponding kernel, $\theta(x, y)$, of the generalized Hilbert transform: (**b**) real and (**c**) imaginary components in spatial domain. (**d-f**) Disk object, radius of 1.0μm and electric susceptibility of 3.0, in (**d**) spatial domain $a(x, y)$ and scattering form factor in **k**-space: $a(k_x, k_y)$ with invisibility (**e**) not activated and (**f**) activated. (**g-l**) Invisibility-activated complex susceptibility profiles after 1, 2 and 3 iterations are depicted in: (**g**),(**h**),(**i**) real and (**j**),(**k**),(**l**) imaginary parts, respectively (the dashed white lines indicate the object contour). The area of invisibility is a Gaussian function (in the form of exp(-($k_x$+2$k_0$)$^2$/(2$\sigma_x^2$)-($k_y$+2$k_0$)$^2$/(2$\sigma_y^2$)) centered at -2*$k_0$=($k_x$, $k_y$)=(-12.6, 0)μm$^{-1}$ with standard deviations of $\sigma_x$=4.0μm$^{-1}$ and $\sigma_y$=8.0μm$^{-1}$ in the $k_x$- and $k_y$- directions, respectively.

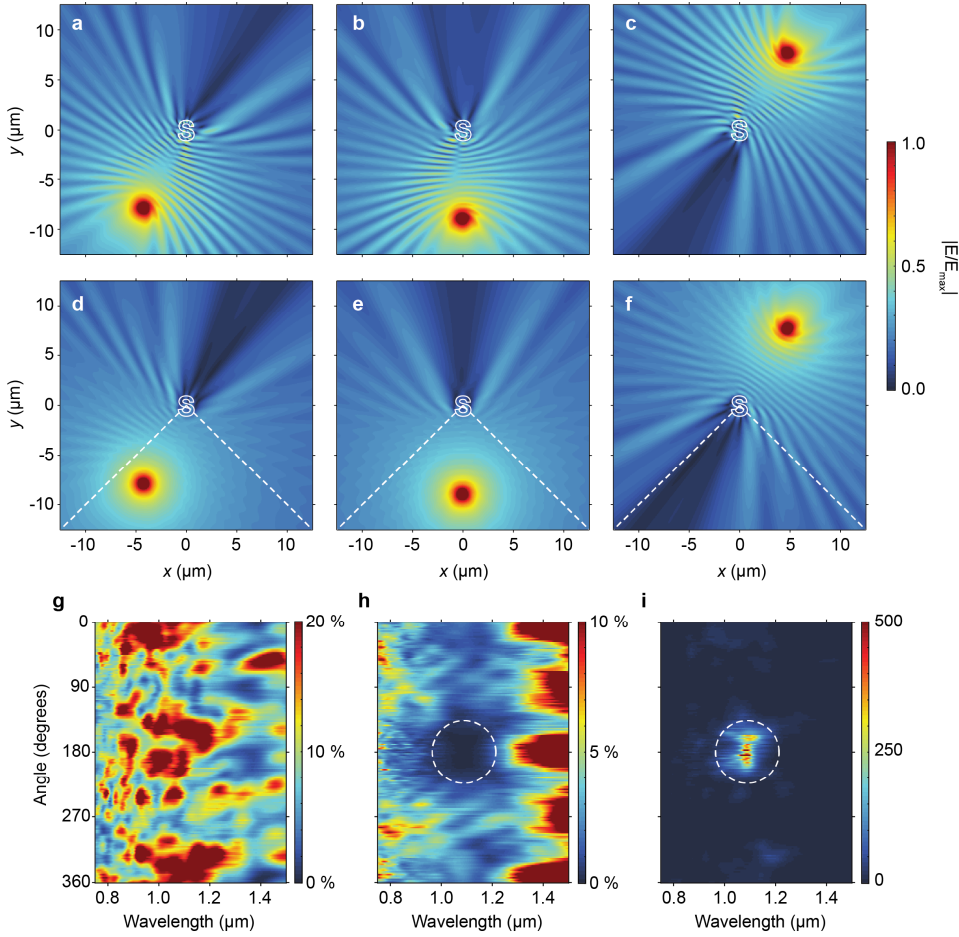

**Figure 3 | Reflection test before and after applying the invisibility on demand.** (**a-f**) FDTD simulations for a "S" shaped object, with an initial susceptibility $\chi=3$, is probed by a monochromatic ($\lambda_0=1.1\mu m$) point source at different angular positions, from left to right: 208°, 180° and 32° (with respect to a vertical axis, in clockwise direction), when: invisibility is (**a**),(**b**),(**c**) not-activated and (**d**),(**e**),(**f**) activated using a Gaussian invisibility function located at $(k_x, k_y)=(0, -20.0)\mu m^{-1}$ with a standard deviation of $7.0\mu m^{-1}$ in both the $k_x$ and $k_y$ directions. The scatterer is outlined by white solid lines. The white dashed lines show the angular range of invisibility on demand. For a full angular analysis, see Supplementary Movie 1. (**g,h**) Angle-wavelength scattering spectra obtained with a plane wave source, for (**g**) invisibility not-activated and (**h**) activated. (**i**), Relative scattering (ratio between scattering spectra for inactivated and activated invisibility) in angle-wavelength domain.

**Invisibility on demand realization.** To numerically verify the "invisibility on demand" proposal, we performed a series of numerical simulations based on the FDTD method[35]. For brevity, we present here the FDTD calculations only for transverse magnetic (TM) polarization (electric field perpendicular to the plane), however the procedure holds also for transverse electric (TE) polarization. Figures 3a,b and 3c show the steady state electric field distribution at the operational wavelength before applying the invisibility procedure for different excitation angles. Similarly, Fig. 3d-f depict the field distributions after applying the invisibility procedure, exactly for the same illumination. It is clear that invisibility is achieved, when comparing the scattering of the object/modified object when illuminated by a probe source at different positions. Whereas scattered waves strongly interfere with incident light, a spatial region of uniform field distribution is a clear sign of invisibility. Moreover in Fig. 3d, a partial standing wave effect outside the invisibility angular range can be observed. This is in particular due to the fact the scattered wavevectors outside the angular range do not lie inside the filtered wavevector region. On the other hand in Fig. 3e, when all the scatterings lie inside the angular range, no standing wave is observed. Furthermore, to reveal the limits of the invisibility region in terms of both the operational wavelength and excitation angle, angle resolved reflection spectra calculations were performed. Figure 3g-h show the cases for which the invisibility is not-activated (R1) and activated (R2), respectively. By comparing Fig. 3g-h, it can be inferred that by applying the invisibility on demand, a strong anti-reflection behavior can be achieved for a specific angle and frequency range. One fact to note is that the overall reflection decreases for nearly all excitation angles after applying the proposed procedure, however the decrease is substantially larger in the designed anti-reflection region. We attribute this fact to the increase of absorption losses. The ratio between R1 and R2, provided in Fig. 3i clearly reveals that the reduction of the reflection is indeed significantly larger within the desired angle and wavelength ranges. A further analysis to quantitatively compare the back-scattered waves inside and outside the designed invisibility angular range has been provided in Supplementary Note 4. Here, we also note the fact that although our design approach is based on the Born approximation, these results suggest than an object with an initial refractive index as high as 2.0 can be rendered invisible if the size of the invisibility areas is correctly chosen. In this regard, Supplementary Note 5 provides a further understanding between the area of the kernel and the obtained invisibility at moderate and large refractive index contrasts.

## Discussion

To conclude, we propose a generalized Hilbert transform relating the two quadratures of the complex susceptibility of an object, to provide invisibility with respect to particular illumination/detection arrangements, and for particular frequency ranges corresponding to arbitrary demands. The invisibility on demand scheme allows either designing objects to be invisible, or alternatively modifying the complex susceptibility of a given object to render invisibility. The scheme succeeds not only in rendering invisible a passively scattering object, characterized by a real susceptibility (with neither gain nor losses), but also for arbitrary complex-valued scattering functions $\chi_{re}(x,y)$. The procedure turns out to be more flexible than the spatial KK transform, while also allowing to form objects with topologically complex shapes at higher dimensions as opposed to previous demonstrations[36,37]. Moreover, following an iterative chain of generalized Hilbert transforms, we propose the invisibility on demand with additional constraints, *i.e.* restricting the required modification of the complex refractive index within practical limits by avoiding gain areas. Finally, we would like to highlight the practical advantages of our findings over transformation optics based invisibility schemes. Namely, as the coordinate transformations in such structures always lead to anisotropic magnetic materials, the optical performance has to be scarified to enable practical realizations. In contrast, the requisite materials in this manuscript are locally isotropic, non-magnetic and gain-free (in the case of bidirectional operation also loss-free). While being feasible in optics, gain materials may result inconvenient, and they remain very challenging for other kind of waves, for instance in acoustics. The procedure is theoretically presented and supported by FDTD numerical simulations of arbitrary-shaped objects.

We present here the basic idea of invisibility on demand, leaving the different aspects of the concept in the supplementary materials: analytically solvable invisibility functions; also special cases leading to broad angle invisibility cloaking, and other more complex cases. The proposed scheme primarily discussed in optics, in principle is working in other fields of wave dynamics.

## Methods

**Numerical calculations.** The FDTD calculations were conducted in a 2D grid (25.00μm x 25.00μm) with a mesh resolution of 0.02μm x 0.02μm. To terminate the computational region, perfectly matched layers were used on all four boundaries. In the case of the field calculations, a monochromatic point probe source was placed at a radial distance of 10.0μm with respect to the center of the simulation region, to illuminate the scatterer from specific angular directions. The illumination angle is defined clockwise from the vertical axis. On the other hand, in the case of reflection spectral calculations, a broadband line source was placed similarly at a distance of 10.00μm. The scatterer was then rotated anti-clockwise to obtain the reflection spectra for all excitation angles with incremental steps equal to $2°$.

**Data availability.** The data that support the findings of this study are available from the corresponding author on request.

(†) Online link to supplementary video 1: http://bit.ly/2nvFbiP and supplementary video 2: http://bit.ly/2nqt0m2

# Acknowledgements

Authors acknowledge financial support of NATO SPS research grant No: 985048, support from Spanish Ministerio de Ciencia e Innovación, and European Union FEDER through project FIS2015-65998-C2-1-P, and partial support of the Turkish Academy of Science.

# Author contributions

K.S. proposed the idea and performed analytical derivations of Hilbert-like transform; Z.H. performed FDTD numerical simulations; M.B. and R.H. developed the iterative procedure to incorporate additional constrains. All authors were involved in discussions, writing and revision of the manuscript.

# Figure Legends

**Figure 1 | Illustration of unidirectional invisibility and invisibility on demand.** (**a**,**b**) In full unidirectional invisibility, in order to prevent left-reflection of every right-propagating wave, (**b**) all modulation components on the left half-plane, $k_x<0$, in $\mathbf{k} = (k_x, k_y)$ space, must be set to zero. **c**,**d**, For invisibility on demand, to prevent back-scattering in a particular angular range (**d**) the modulation components to be set to zero are just the ones within a limited invisibility area in $\mathbf{k} = (k_x, k_y)$ space. In particular, the invisibility case (**c**) requires uncoupling the scattering of incident waves around $\mathbf{k} = (k_0, 0)$ into back reflections, i.e. into waves in the vicinity of $\mathbf{k} = (-k_0, 0)$; then, the invisibility function has to be centered at $\mathbf{k} = (-2k_0, 0)$.

**Figure 2 | Invisibility on demand procedure workflow in spatial and wavevector domain.** (**a-c**) Area of invisibility in (**a**) **k**-space: $\theta(k_x, k_y)$ and the corresponding kernel, $\theta(x, y)$, of the generalized Hilbert transform: (**b**) real and (**c**) imaginary components in spatial domain. (**d-f**) Disk object, radius of 1.0μm and electric susceptibility of 3.0, in (**d**) spatial domain $a(x, y)$ and scattering form factor in **k**-space: $a(k_x, k_y)$ with invisibility (**e**) not activated and (**f**) activated. (**g-l**) Invisibility-activated complex susceptibility profiles after 1, 2 and 3 iterations are depicted in: (**g**),(**h**),(**i**) real and (**j**),(**k**),(**l**) imaginary parts, respectively (the dashed white lines indicate the object contour). The area of invisibility is a Gaussian function (in the form of exp(-$(k_x+2k_0)^2/(2\sigma_x^2)$-$(k_y+2k_0)^2/(2\sigma_y^2)$)) centered at -2*$k_0$=($k_x$, $k_y$)=(-12.6, 0)μm$^{-1}$ with standard deviations of $\sigma_x$=4.0μm$^{-1}$ and $\sigma_y$=8.0μm$^{-1}$ in the $k_x$- and $k_y$- directions, respectively.

**Figure 3 | Reflection test before and after applying the invisibility on demand.** (**a-f**) FDTD simulations for a "S" shaped object, with an initial susceptibility $\chi$=3, is probed by a monochromatic ($\lambda_0$=1.1μm) point source at different angular positions, from left to right: 208°, 180° and 32° (with respect to a vertical axis, in clockwise direction), when: invisibility is (**a**),(**b**),(**c**) not-activated and (**d**),(**e**),(**f**) activated using a Gaussian invisibility function located at ($k_x$, $k_y$)=(0, -20.0)μm$^{-1}$ with a standard deviation of 7.0μm$^{-1}$ in both the $k_x$ and $k_y$ directions. The scatterer is outlined by white solid lines. The white dashed lines show the angular range of invisibility on demand. For a full angular analysis, see Supplementary Movie 1. (**g**,**h**) Angle-wavelength scattering spectra obtained with a plane wave source, for (**g**) invisibility not-activated and (**h**) activated. (**i**), Relative scattering (ratio between scattering spectra for inactivated and activated invisibility) in angle-wavelength domain.

# Supplementary information - Invisibility on demand based on a generalized Hilbert transform

## Supplementary Note 1: Analytical kernels

Some invisibility functions, or different shapes of the invisibility area, allow obtaining analytical expressions for the kernels of the generalized Hilbert transform. Here we list some of them.

The kernel of an elliptical (rectangular) invisibility area, centered around the wavevector of the probe wave is expressed in Bessel (Sinc) functions:

$$\theta(k_x, k_y) = \mathrm{Disc}\left(\sqrt{\left(\frac{k_x + 2k_0}{w_x}\right)^2 + \left(\frac{k_y}{w_y}\right)^2}\right) \to \theta(x,y) = w_x w_y\, e^{\mathrm{i}\frac{2k_0 x}{w_x}} \frac{J_1(\sqrt{w_x^2 x^2 + w_y^2 y^2})}{\sqrt{w_x^2 x^2 + w_y^2 y^2}} \quad (S1)$$

$$\left(\theta(k_x, k_y) = \mathrm{Rect}\left(\frac{k_x + 2k_0}{w_x}\right)\mathrm{Rect}\left(\frac{k_y}{w_y}\right) \to \theta(x,y) = w_x w_y\, e^{\mathrm{i}\frac{2k_0 x}{w_x}} \mathrm{Sinc}(w_x x)\mathrm{Sinc}(w_y y)\right) \quad (S2)$$

These cases are presented in Supplementary Fig. 1a-c (d-f).

The invisibility area in form of a ring corresponding to all possible scattered wavevectors of fixed frequency, is of a special interest, since it allows full invisibility with respect to an incident monochromatic plane wave, in a given direction. The difference of two *Disc* functions with different radii $k_0(1 + w)$ and $k_0(1 - w)$, results in a ring with a width, $2wk_0$:

$$\theta(k_x, k_y) = \mathrm{Disc}\left(\sqrt{1-w}\sqrt{\left(\frac{k_x}{k_0}+1\right)^2 + \left(\frac{k_y}{k_0}\right)^2}\right) - \mathrm{Disc}\left(\sqrt{1+w}\sqrt{\left(\frac{k_x}{k_0}+1\right)^2 + \left(\frac{k_y}{k_0}\right)^2}\right)$$

$$\to \theta(x,y) = \frac{k_0^2}{1-w} e^{\mathrm{i}x\sqrt{1-w}} \frac{J_1\left(\sqrt{\frac{k_0^2}{1-w}(x^2+y^2)}\right)}{\sqrt{\frac{k_0^2}{1-w}(x^2+y^2)}} - \frac{k_0^2}{1+w} e^{\mathrm{i}x\sqrt{1+w}} \frac{J_1\left(\sqrt{\frac{k_0^2}{1+w}(x^2+y^2)}\right)}{\sqrt{\frac{k_0^2}{1+w}(x^2+y^2)}} \quad (S3)$$

As it is represented in Supplementary Fig. 1g-i.

Another interesting case is the "solar eclipse" shaped invisibility area given by the difference of two *Disc* functions with different radii touching at the origin **k** = 0, in **k**-space. This case allows covering all possible scattered wavevectors, keeping the incident monochromatic plane wave unaffected, obtaining the unidirectional cloaking effect. The width of the invisible area changes with the scattering angle, and becomes maximum at normal reflection. The analytical expression of this invisibility areas and corresponding kernel are:

$$\theta(k_x, k_y) = \mathrm{Disc}\left(\sqrt{\left(\frac{k_x}{k_0(1+w)}+1\right)^2 + \left(\frac{k_y}{k_0(1+w)}\right)^2}\right) - \mathrm{Disc}\left(\sqrt{\left(\frac{k_x}{k_0(1-w)}+1\right)^2 + \left(\frac{k_y}{k_0(1-w)}\right)^2}\right)$$

$$\to \theta(x,y) = k_0 e^{\mathrm{i}x} \frac{(1+w)J_1\left(\sqrt{k_0(1+w)(x^2+y^2)}\right) - (1-w)J_1\left(\sqrt{k_0(1-w)(x^2+y^2)}\right)}{\sqrt{(x^2+y^2)}}$$

(S4)

Note that these last two cases, Supplementary Figs. 1g-i and 1j-l, lead to kernels which are weaker in amplitude, but more delocalized in space.

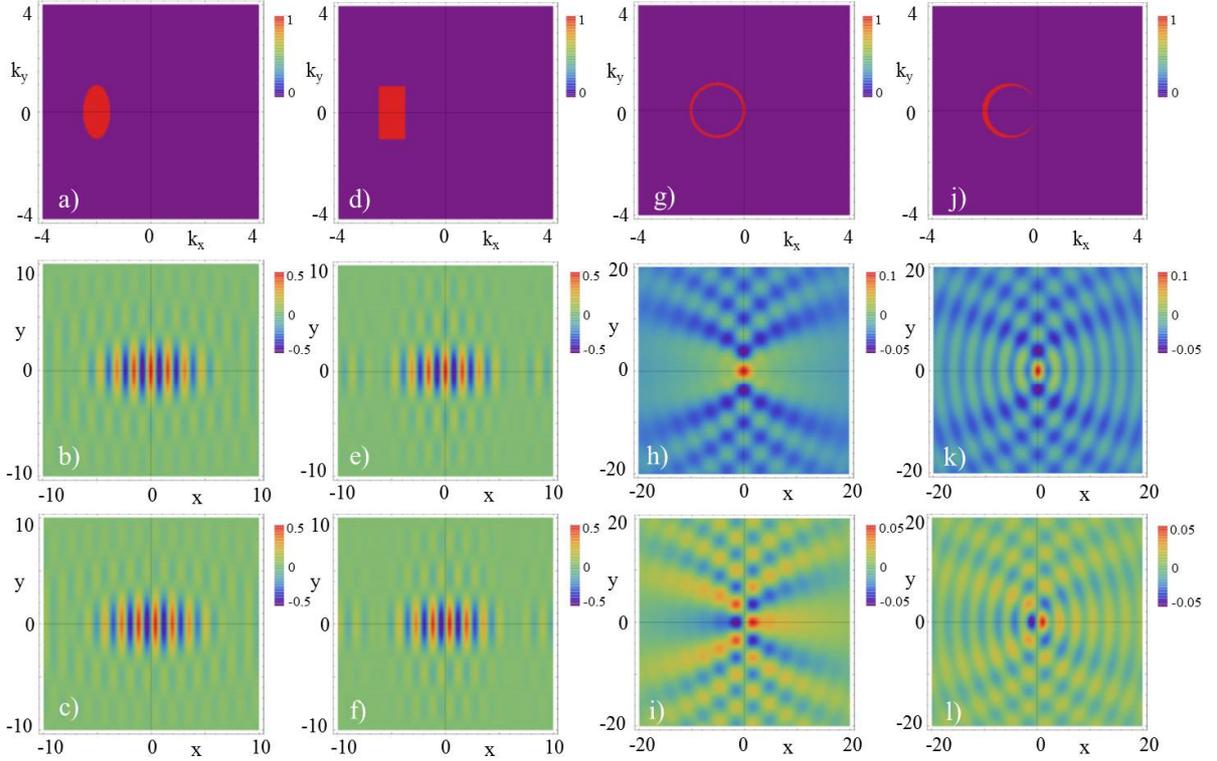

**Supplementary Figure 1 | Invisibility areas in Fourier space and corresponding kernels**. Invisibility areas with: (**a**) elliptical shape (first column), (**d**) rectangular shape (second column) both with widths $w_x=0.5k_0$ and $w_y=1.0k_0$ —where $k_0$ is the input wavenumber of light— [which corresponds to 30º]; (**g**) complete circle shape (third column) and (**j**) "solar eclipse" shaped (fourth column) for these last two $w=0.1\lambda_0$. Corresponding associated kernels of the generalized Hilbert transform, (**b**),(**e**),(**h**),(**k**) real parts and (**c**),(**f**),(**i**),(**l**) imaginary parts.

## Supplementary Note 2: Analytical calculation of invisibility on demand

This supplementary provides the calculation of the invisibility on demand procedure for a simple 1D or 2D Gaussian object (with either real or/and imaginary parts of the susceptibility) with a Gaussian invisibility function. In this case the analytical expressions are derived, which allow estimating the parameters of the system. In particular, it provides an estimation on the strength of the required modification of the potential to obtain invisibility on demand.

First, we consider a 1D Gaussian object of width $x_0$:

$$a(x) = a_0 \exp(-x^2/x_0^2) \tag{S4}$$

with corresponding spectrum (Fourier transform),

$$a(k) = \frac{1}{\sqrt{2\pi}} \int a(x) \exp(-ikx) dx = \frac{x_0 a_0}{\sqrt{2}} \exp(-k^2 x_0^2/4). \tag{S5}$$

and we consider a Gaussian invisibility function:

$$\theta(k) = \exp(-(k+2k_0)^2/k_{\text{inv}}^2) \tag{S6}$$

with halfwidth $k_{\text{inv}}$; being $k_0 = \omega/c$ the wavevector of the object illumination (the invisibility function must be centered at around $-2k_0$ in order to prevent scattering from wavevector $k_0$ to $-k_0$).

The kernel of the generalized Hilbert transform is:

$$\theta(x) = \frac{1}{\sqrt{2\pi}} \int \theta(k) \exp(ikx) dx = \frac{\sqrt{2}}{x_{\text{inv}}} \exp(-2ik_0 x - x^2/x_{\text{inv}}^2). \tag{S7}$$

Here $x_{\text{inv}}^2 = 4/k_{\text{inv}}^2$

According to the procedure described in the main article, the modification of the object is the corresponding convolution of the object susceptibility profile with the invisibility function. The calculation of such convolution in **k**-space results in:

$$a_1(x) = a(x) - \frac{1}{\sqrt{2\pi}} \int a(k) \theta(k) \exp(ikx) dk \tag{S8}$$

Inserting the Gaussian functions, and after some algebra, it follows that:

$$a_1(x) = a(x) - \frac{a_0 x_0}{x_1} \exp\left(-ik_{c,1} x - \frac{x^2}{x_1^2} - \frac{x_0^2 x_{\text{inv}}^2 k_0^2}{x_1^2}\right) \tag{S9}$$

The modification of the object results to be again a Gaussian function, centered at:

$$k_{c,1} = \frac{-2k_0 x_{\text{inv}}^2}{x_1^2} \tag{S10}$$

in wavenumber domain, its width and amplitude being:

$$x_1^2 = x_0^2 + x_{\text{inv}}^2; \qquad \frac{a_0 x_0}{x_1} exp\left(-\frac{x_0^2 x_{\text{inv}}^2 k_0^2}{x_1^2}\right) \tag{S11}$$

Note that the modification of the object is a Gaussian function contains the oscillatory multiplier: $\exp(-ik_{c,1}x)$. This indicates that the invisibility on demand introduces a modulation on the real and imaginary parts of susceptibility.

The relative norm of the correction (with respect to the norm of the initial object) is:

$$\exp\left(-\frac{x_0^2 x_{\text{inv}}^2 k_0^2}{x_1^2}\right) \tag{S12}$$

which exponentially decreases with the decreasing area of invisibility in k-space.

The derived expression is valid for either an object with a Gaussian shape entailing only losses (being $a_0$ a real number) as well as for a Gaussian index profile (being $a_0$ imaginary), or for both simultaneously, i.e. for a complex index Gaussian profile.

In a limit of a small invisibility area in **k**-space, $x_{\text{inv}}^2 \gg x_0^2$, therefore assuming $x_1^2 \approx x_{\text{inv}}^2$, the normalized correction function is:

$$\frac{\Delta a(x)}{a_0} \approx \frac{x_0}{x_{\text{inv}}} exp\left(-2ik_0 x - \frac{x^2}{x_{\text{inv}}^2} - x_0^2 k_0^2\right) \tag{S13}$$

i.e. it is weak in amplitude, but broad in space; also, the modulation is centered at wavenumber $-k_{c,1} \approx -2ik_0$. In the opposite limit, of a broad invisibility area in **k**-space, $x_{\text{inv}}^2 \ll x_0^2$, and $x_1^2 \approx x_0^2$, leads to the correction function:

$$\frac{\Delta a(x)}{a_0} \approx \exp\left(-ik_c x - \frac{x^2}{x_0^2} - x_{\text{inv}}^2 k_0^2\right) \tag{S14a}$$

$$k_{c,1} = \frac{-2k_0 x_{\text{inv}}^2}{x_0^2} \tag{S14b}$$

which is strong in amplitude (on the same order as the object itself), but narrow in space (nearly of the same width as the object); the modulation is centered at wavenumber $-k_{c,1} \approx 0$. However, in all cases the amplitude of the correction term is always smaller than the amplitude as the object itself.

As the correction of the scattering object ($a_0$ is real-valued) contains gain and loss (due to oscillatory factor $exp(-ik_{c,1}x)$ in (S9), perhaps the simplest way to regularize the object (to remove the gain areas from a modified object in (S9) is to add the corresponding loss profile:

$$a_1(x) = a(x) - \frac{a_0 x_0}{x_1}\exp\left(-ik_c x - \frac{x^2}{x_1^2} - \frac{x_0^2 x_{inv}^2 k_0^2}{x_1^2}\right) + i\frac{a_0 x_0}{x_1}\exp\left(-\frac{x^2}{x_1^2} - \frac{x_0^2 x_{inv}^2 k_0^2}{x_1^2}\right) \quad (S15)$$

Such added loss profile will result in a weak scattering into invisibility domain (will spoil the invisibility achieved by (S9). This, however, can be removed again by using the same generalized Hilbert transform acting on (S15) rendering the object invisible again, by the next correction order:

$$\Delta a_1(x) = -\frac{a_0 x_0^2}{x_1 x_2}\exp\left(-ik_{c,2}x - \frac{x^2}{x_2^2} - \frac{x_0^2 x_{inv}^2 k_0^2}{x_2^2}\right) \quad (S16)$$

with the parameters (half-width, and center wavenumber):

$$x_2^2 = x_0^2 + 2x_{inv}^2; \qquad k_{c,2} = \frac{-2k_0 x_{inv}^2}{x_2^2} \quad (S17)$$

The procedure can be repeated, leading to a converging series. The convergence is assured since for each $x_n^2 = x_0^2 + nx_{inv}^2$ ; $x_n^2 > x_0^2$, and $exp(-x_0^2 x_{inv}^2 k_0^2/x_2^2)<1$. This proves that the invisibility on demand can always be obtained without gain for a Gaussian-shaped object, with a Gaussian invisibility function. This also hints how to construct and invisible on demand object without gain for arbitrary shaped object, and with an arbitrary invisibility function, however it does not provide a rigid proof.

The 1D Gaussian case can be directly extended to 2D, for elliptic areas of invisibility:

$$a(x,y) = a_0\exp(-x^2/x_0^2 - y^2/y_0^2) \quad (S18)$$

$$\theta(k_x, k_y) = \exp(-(k_x + 2k_0)^2/k_{x,inv}^2 - k_y^2/k_{y,inv}^2) \quad (S19)$$

Inserting the Gaussian functions, and after some algebra, it follows that:

$$a_1(x,y) = a(x,y) - \frac{a_0 x_0 y_0}{x_1 y_1}exp\left(-ik_{c,1}x - \frac{x^2}{x_1^2} - \frac{y^2}{y_1^2} - \frac{x_0^2 x_{inv}^2 k_0^2}{x_1^2}\right) \quad (S20)$$

The procedure of regularization of (S20) is analogous to that in 1D case.

## Supplementary Note 3: Invisibility on demand procedure applied on complex shaped objects

To evidence that the proposed procedure holds also for complex shaped objects, we applied the invisibility on demand procedure for an "Einstein face" shaped object (with initial susceptibility of 2.3, see Supplementary Fig. 2), and considering a Gaussian kernel placed at $(k_x, k_y)=(0, -20.0)µm^{-1}$ and a standard deviation of $7.0µm^{-1}$ in both the $k_x$ and $k_y$ directions.

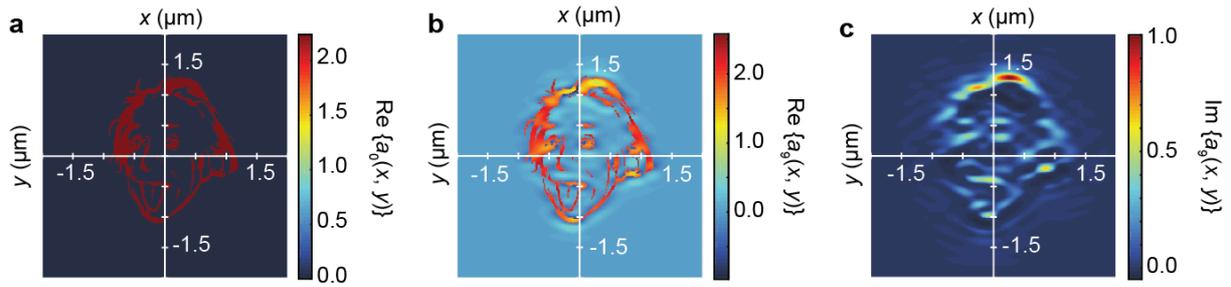

**Supplementary Figure 2 | Invisibility on demand procedure applied on a complex shaped object.** (**a-c**) Complex susceptibility profiles of an "Einstein face" shaped object modified with a Gaussian invisibility area located at $(k_x, k_y)=(0, -20.0)\mu m^{-1}$ with a standard deviation of $7.0\mu m^{-1}$ in both the $k_x$ and $k_y$ directions. (**a**) The real part of the initial object and the (**b**) real and (**c**) imaginary parts of the modified complex susceptibility after 9 iterations.

The FDTD simulations before and after applying the invisibility procedure are given in Supplementary Fig. 3 at the operational wavelength of 1.10μm. It can be inferred from the figure that within the targeted angle range (delimited with white dashed lines) reflections are significantly suppressed, as the interference patterns between the probe source and reflections from the object are strongly suppressed.

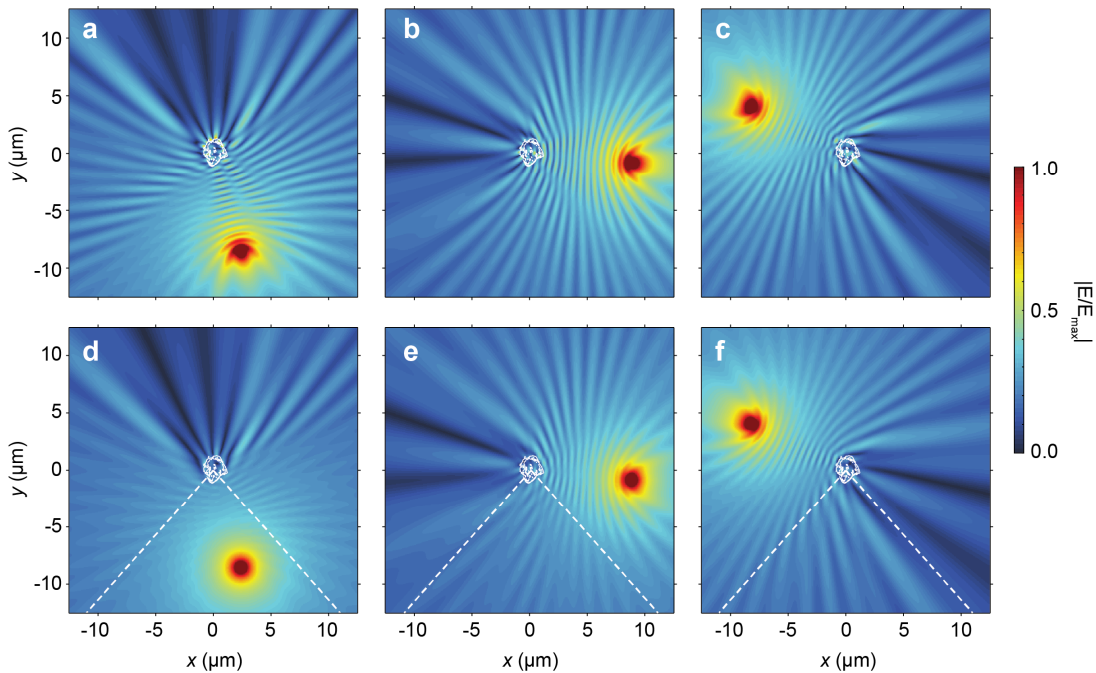

**Supplementary Figure 3 | Numerical results of a complex shaped scatterer.** (**a-f**) FDTD simulations of waves coming from point sources at different positions. Scattering object is an "Einstein face" shaped object given in Supplementary Fig. 2. The upper and lower rows correspond to cases where the invisibility is not-activated and activated, respectively. The white dashed lines denote the angle range where invisibility is activated. For a full angular analysis, see Supplementary Movie 2.

Another intriguing kernel type is the "solar eclipse" shaped invisibility area (see Supplementary Note 1), which yields elimination of reflection in a large angle. Here, we modify the "Einstein face" shaped object by calculating the generalized Hilbert transform with a kernel given by Eq. (S4). Supplementary Figure 4 shows the complex susceptibility profiles of the original and the modified object.

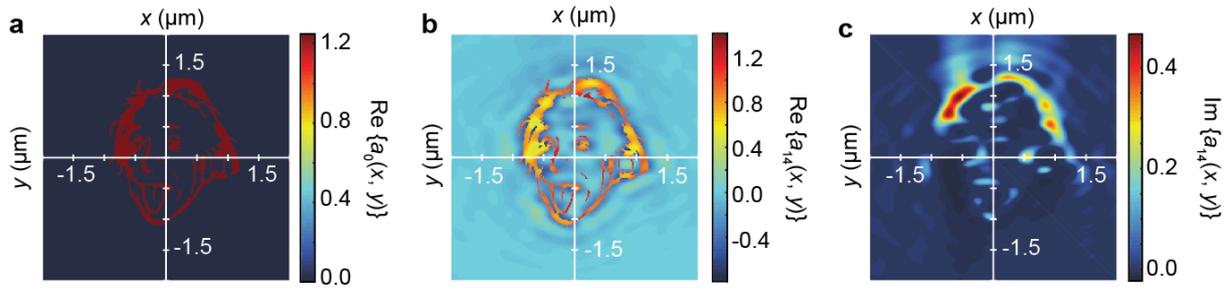

**Supplementary Figure 4 | Invisibility on demand procedure applied with a "solar eclipse" shaped kernel.** (**a-c**) Complex susceptibility profiles of an "Einstein face" shaped object modified with a "solar eclipse" shaped invisibility area. (**a**) The real part of the initial object and the (**b**) real and (**c**) imaginary parts of the modified complex susceptibility after 14 iterations. The inner and outer disks of the "solar eclipse" shaped invisibility are located at $(k_x, k_y)=(0, -8.0)\mu m^{-1}$, $(0, -12.0)\mu m^{-1}$ and have radii equal to $8.0\mu m^{-1}$, $12.0\mu m^{-1}$; respectively.

Supplementary Fig. 5 shows the FDTD results for the original and the modified object. It is evident that for a given incidence angle (denoted by a white dashed line) the reflections disappear at a large angular range, as can be judged from the disappearance of the fringes. Only the scattering at small deflection angles remains.

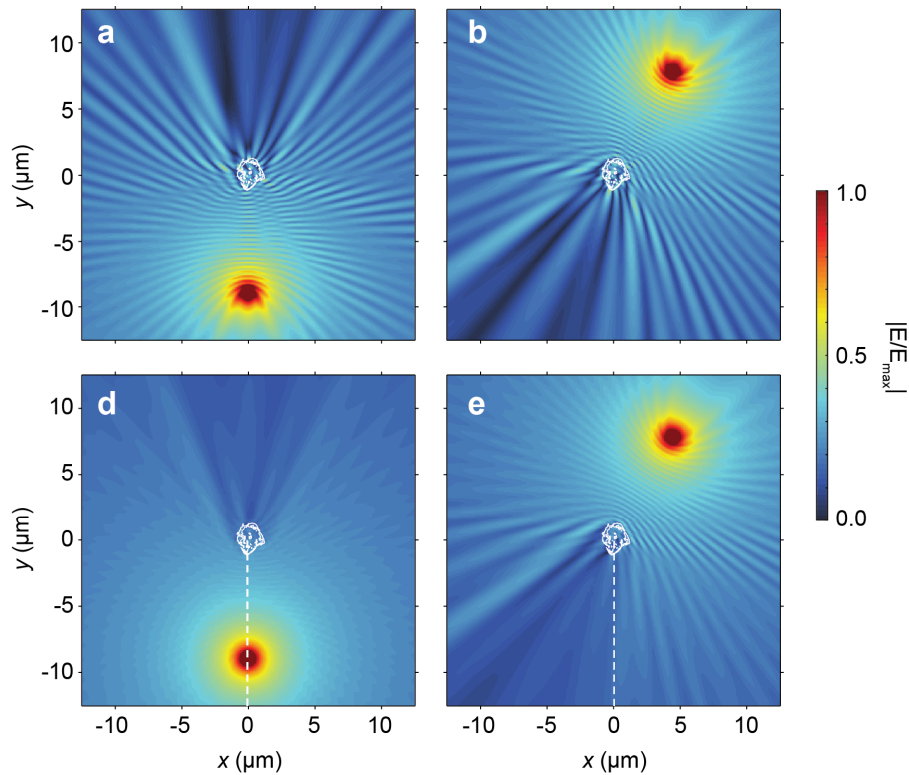

**Supplementary Figure 5 | Numerical results for the case of a "solar eclipse" shaped kernel.** (**a-d**) FDTD simulations of waves coming from point sources at different positions. Scattering object is an "Einstein face" shaped object given in Supplementary Fig. 4. The upper and lower rows correspond to cases where the invisibility is not-activated and activated, respectively. The white dashed lines denote the incident angle where invisibility is activated. The operational wavelength corresponds to $0.7\mu m$.

One interesting property of the proposed approach is that for symmetrically placed invisibility areas (that is to say for bidirectional invisibility at a specific angle and frequency range), the need for a gain loss modulation is completely eliminated (since the Fourier transform of an even function is purely real). Supplementary Fig. 6 show the complex susceptibility profiles before

(Supplementary Fig. 6a) and after (Supplementary Fig. 6b, c) applying the invisibility procedure for the same "S" shaped object as in Fig. 3 (in the manuscript). As can be noted from this figure, the modified object is composed of a purely real dielectric susceptibility. Furthermore, Supplementary Fig. 7 shows the FDTD results for such a case, where the suppression of the reflections for both symmetrically placed angle ranges (delimited with white dashed lines) can be observed.

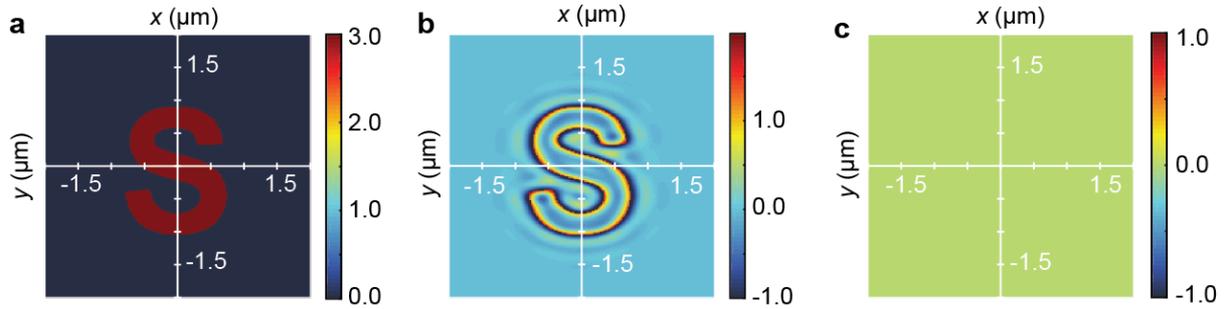

**Supplementary Figure 6 | Invisibility on demand procedure applied with two symmetrically shaped kernels.** (**a-c**) Complex susceptibility profiles of an "S" shaped object modified with two symmetrically shaped invisibility areas. (**a**) The real part of the initial object and the (**b**) real and (**c**) imaginary parts of the modified complex susceptibility. The invisibility areas are located at $(k_x, k_y)=(0, \pm11.4)\mu m^{-1}$ with a standard deviation of $6.0\mu m^{-1}$ in both the $k_x$ and $k_y$ directions.

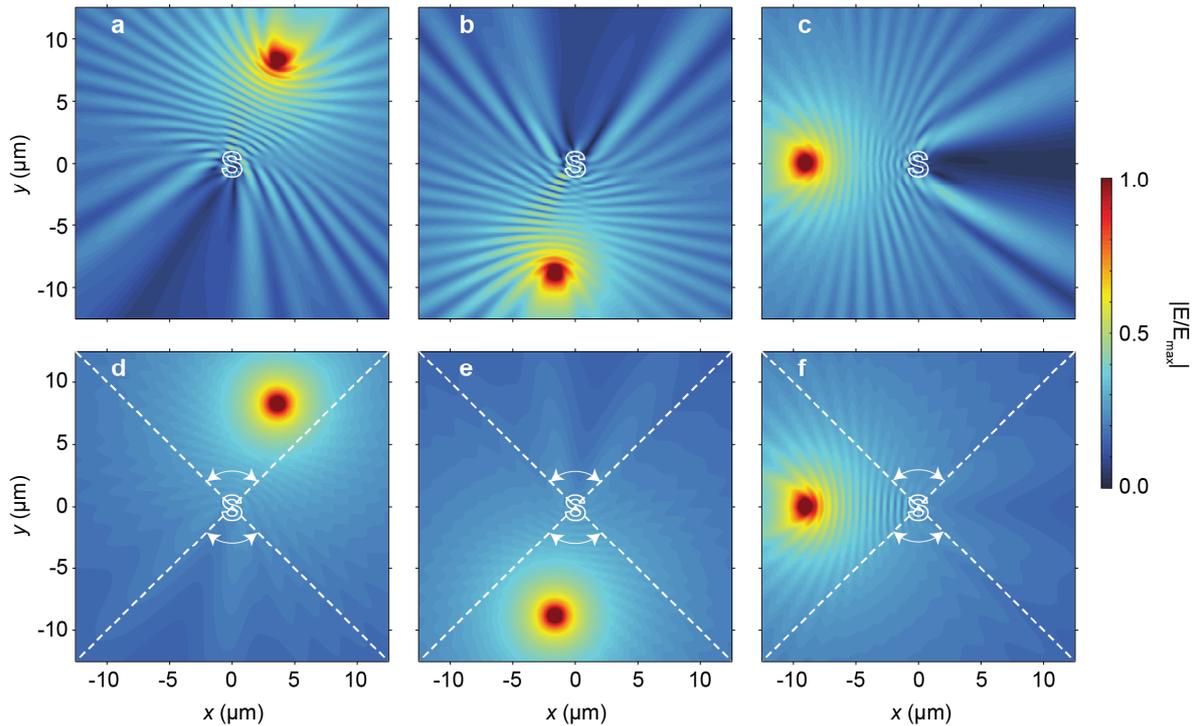

**Supplementary Figure 7 | Reflection test for two symmetrically placed kernels.** (**a-f**) FDTD simulations of waves coming from point sources at different positions for scatterer given in Supplementary Fig. 6. The upper and lower rows correspond to cases where the invisibility is not-activated and activated, respectively. Excitation angles correspond to (**a,d**) 24°, (**b,e**) 190° and (**c,f**) 270°. The white dashed lines denote the angular range where invisibility is activated. Operational wavelength corresponds to 1.1μm.

# Supplementary Note 4: Examination of invisibility angular range

To further verify that the back-scattering is suppressed in the designed angular range, we have calculated the standing wave ratio (SWR) for various illumination angles. The SWR values have been determined by taking the ratio of the maximum magnitude of the electric field to its minimum value under plane wave illumination. Supplementary Figs. 8a (invisibility not activated) and 8b (invisibility activated) show the scattering potential for the same "S" shaped object as in Fig. 3 (in the manuscript), while Supplementary Figs. 8c (invisibility not activated) and 6d (invisibility activated) show the cross sectional field profiles under plane wave illumination.

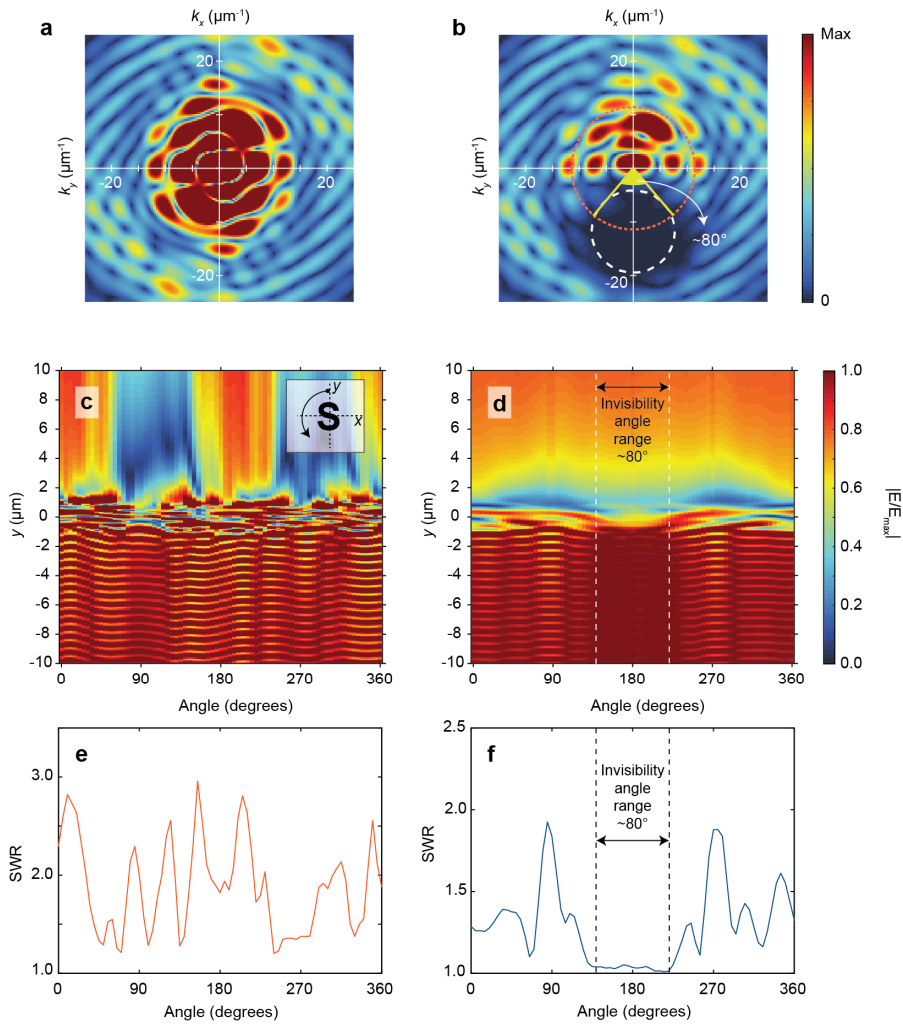

**Supplementary Figure 8 | Examination of the invisibility angular range.** (**a,b**) Absolute value of the complex potential for objects with invisibility (**a**) not-activated and (**b**) activated. (**c,d**) Magnitude of the electric field along the propagation $y$- direction at $x=0$ for the invisibility (**c**) not-activated and (**d**) activated, by rotating the scatterer counter-clockwive, therefore changing the excitation angle (see upper right inset in (**c**)). (**e,f**) The SWR values for the invisibility (**e**) not-activated and (**f**) activated cases. The mean SWR values inside the invisibility range for the invisibility not-activated and activated cases have been determined as 2.16 and 1.03, respectively. Object is the same "S" shaped object as in Fig. 3 (in the manuscript) having an initial susceptibility $\chi=3$. Invisibility area is a Gaussian function (denoted with white dashed lines in (**b**)) located at $(k_x, k_y)=(0, -11.4)$µm$^{-1}$ with a standard deviation of 6.0µm$^{-1}$ in both the $k_x$ and $k_y$ directions. In (**b**) the solid yellow lines delimit the angular range of the invisibility, while the dashed orange line delimit the region formed by the constant wavevector circles having incident angles from 0 to 360°. Light propagation is in the $+y$ direction and operational wavelength is equal to 1.1µm.

The deduced SWR values are provided in Supplementary Figs. 8e (invisibility not activated) and S8f (invisibility activated). The designed angular range of invisibility (see Supplementary Fig. 8b) corresponds to approximately 80°. It is clear from Supplementary Figs. 8d and 8f that the numerically observed angular range matches well with this value. Here, another point worth to mention is that the SWR values outside the angular range are also slightly modified in the invisibility activated case, as was also observed in Fig. 3h (in the manuscript). Similarly, we attribute this observance to the additional absorption losses and to the overall modification of the scattering properties (in all directions) by the filtered wavevector region.

## Supplementary Note 5: Relationship between invisibility performance and the kernel

In order to inspect the intrinsic relation between the performance of invisibility and the kernel, we performed a series of calculations on the "S" object (given in Fig. 3) with varying sizes of elliptically shaped kernels. Supplementary Figure 9a shows the standing wave ratio (SWR) variation for elliptically shaped invisibility kernels for a relatively high index contrast of 2. The back-scattered waves are significantly reduced when the standard deviation is larger than $2\mu m^{-1}$ in both directions, and it is almost suppressed for deviations larger than $4\mu m^{-1}$. Moreover, Supplementary Fig. 9a further proves that even when the index contrast is increased up to 3, higher-order back-scattering can be still suppressed for sufficiently large kernels. These results suggest that the proposed method may attain full invisibility in the desired directions with a careful design of the kernel size and shape, even for higher refractive indices.

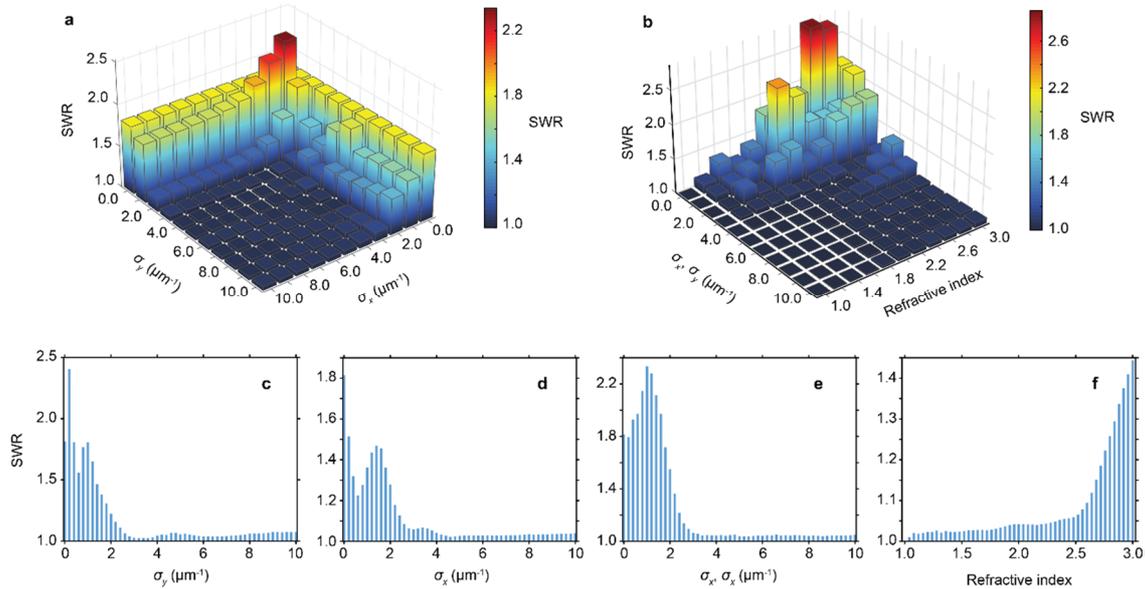

**Supplementary Figure 9 | Invisibility dependence on the kernel size and index contrast.** (**a**,**b**) SWR values for (**a**) elliptically and (**b**) circularly shaped invisibility kernels. The object is the same "S" shaped object as in Fig. 3 (of the manuscript), with an index contrast of 2.0 in (**a**) and for increasing values up to 3 in (b). The invisibility kernel is a Gaussian function located at $(k_x, k_y)=(0, -11.4)\mu m^{-1}$, while having standard deviations of $\sigma_x$ and $\sigma_y$ in the $k_x$ and $k_y$ directions, respectively. (**c-f**) Cross sectional SWR profiles for (**c**) $\sigma_x=4.0\mu m^{-1}$ and initial index of 2.0 (**d**) $\sigma_y=4.0\mu m^{-1}$ and initial index of 2.0 (**e**) initial index of 2.0 and (**f**) $\sigma_x=\sigma_y=4.0\mu m^{-1}$. The operational wavelength is equal to 1.1μm and the wave propagation is in the $+y$ direction.

Next, we show how higher-order scattered waves may be efficiently suppressed by carefully designing the kernel, as for instance for the "solar eclipse" shaped invisibility region. To illustrate this, it is sufficient to inspect the *n*th term in the scattering series:

$$e_s^{(n)}(k) = \frac{-k_0^2 G(k)}{2\pi} \int a(k - k') \tilde{e}_s^{(n-1)}(k') dk' \quad (S21)$$

where $e_s^{(n)}$ is the nth-order scattered electric field, $k_0$ is the free space wavevector, $G(k) = (k_0^2 - k^2)^{-1}$ is the Fourier spectrum of retarded Green function, $a(k)$ and $\tilde{e}_s(k)$ are the Fourier transforms of the complex potential and the scattered field, respectively. It can be directly seen from this equation that when $e_s^{(1)}$ (first-order Born approximation) is completely suppressed, then every successive order will be also zero, regardless of the value of other terms. Another way to understand this is to think that the eclipse actually eliminates all primary scattered propagating waves with respect to a particular plane wave of incidence. The secondary scattering, which is the dominating part of the second Born approximation then disappear as well, since the primarily scattered propagating waves are absent.

To illustrate this full invisibility, we employ the same "S" shaped object as in the above analyses and modify its complex susceptibility with a "solar eclipse" shaped kernel. Supplementary Fig. 10a-c shows the initial and the modified complex susceptibility profiles and Supplementary Fig. 11a-i depicts the corresponding spatial field distributions under point source excitation for high initial refractive index contrasts of 2.0, 3.0 and 4.0. As it follows from these figures, scattering is almost fully suppressed in all directions and almost vanishes even for high index contrasts.

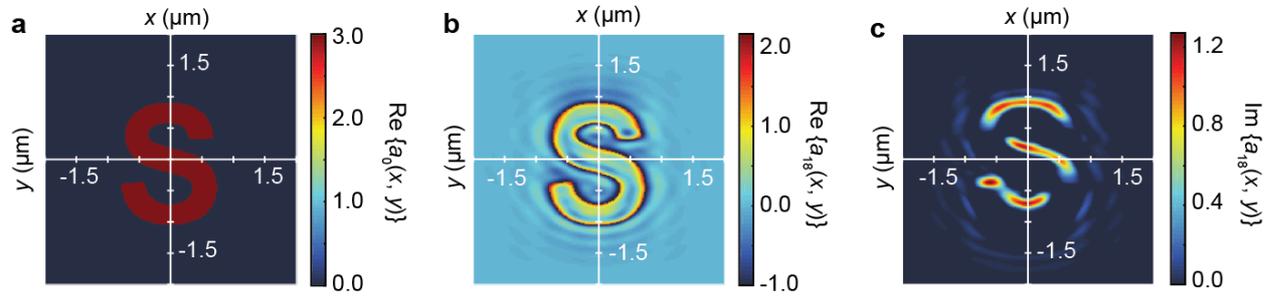

**Supplementary Figure 10 | Invisibility on demand procedure applied with a "solar eclipse" shaped kernel.** (**a-c**) Complex susceptibility profiles of an "S" shaped object modified with a "solar eclipse" shaped invisibility area. (**a**) The real part of the initial object and the (**b**) real and (**c**) imaginary parts of the modified complex susceptibility after 18 iterations. The inner and outer disks of the "solar eclipse" shaped invisibility area are located at $(k_x, k_y)=(0, -5.5)\mu m^{-1}$, $(0, -13.5)\mu m^{-1}$ and have radii equal to $5.5\mu m^{-1}$, $13.5\mu m^{-1}$; respectively.

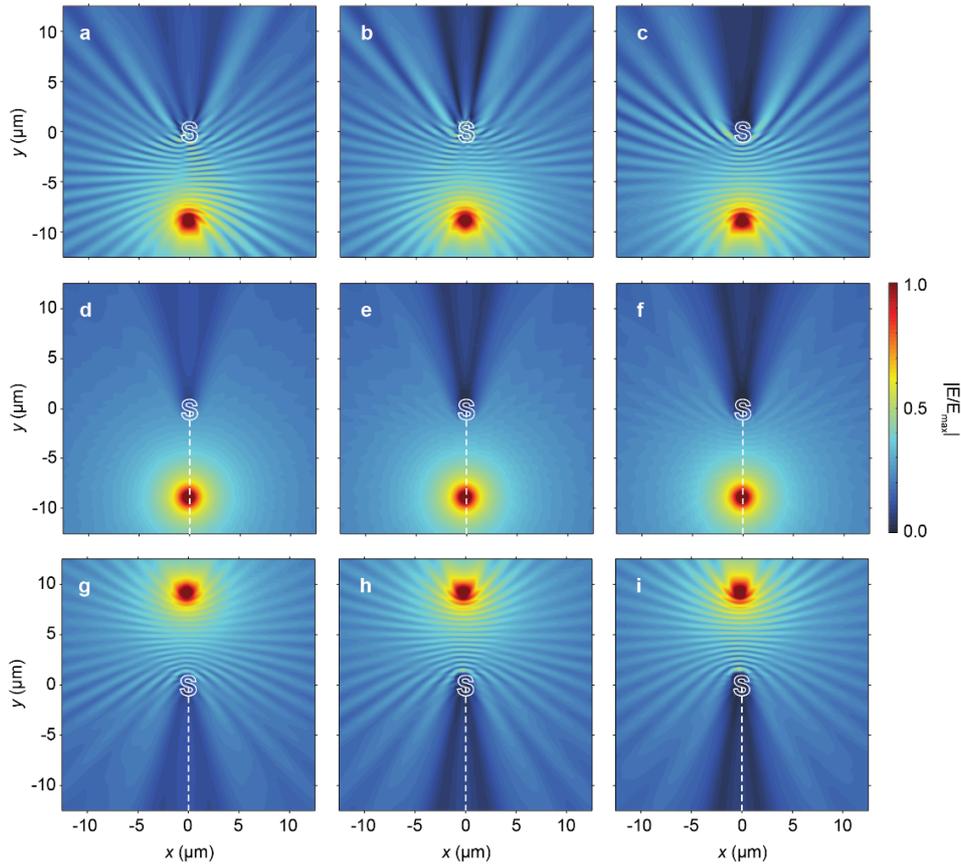

**Supplementary Figure 11 | Numerical results for the case of a "solar eclipse" shaped kernel with high refractive index contrasts.** (**a-i**), FDTD simulations of waves coming from point sources at different positions towards the scatterer presented in Supplementary Fig. 10 with an initial refractive index of (**a,d,g**) 2.0, (**b,e,h**) 3.0 and (**c,f,i**) 4.0. The upper row corresponds to the cases where the invisibility is not activated, the middle row corresponds to invisibility activated cases, and the lower row corresponds to cases where the invisibility is activated but where the source is located outside the invisibility range. The white dashed lines denote the incident angle where invisibility is activated. The operational wavelength is equal to 1.0μm.

To quantify the invisibility performance we evaluate the SWR for various initial refractive indices, the results are provided in Supplementary Fig. 12. The SWR values stay very low and do not experience any significant increase even for high indices, precisely due to the absence of first-order scattered waves as above explained above.

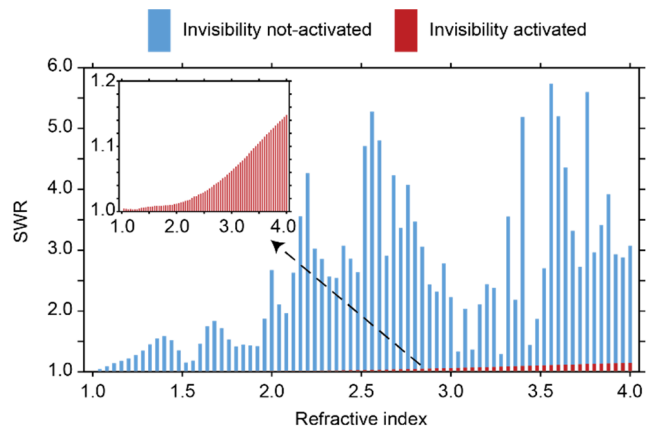

**Supplementary Figure 12 | Invisibility dependence on refractive index contrast for a "solar eclipse" shaped invisibility area.** Superimposed SWR values for the invisibility activated and not-activated cases, for various initial refractive indices of the "S" shaped object presented in Supplementary Fig. 11. The operational wavelength is equal to 1.0μm.